# Inverse design of optical elements based on arrays of dielectric spheres

ALAN ZHAN[1], TAYLOR K. FRYETT[2], SHANE COLBURN[2], ARKA MAJUMDAR[1,2,*]

[1]Department of Physics, University of Washington, Seattle, WA-98195
[2]Department of Electrical Engineering, University of Washington Seattle, WA-98195
*Corresponding author: arka@uw.edu

**Arrays of wavelength scale scatterers are a promising platform for designing optical elements with a compact footprint. The large number of degrees of freedom in this system allows for unique and plentiful functionalities. However, the many variables also create a complex design problem. While intuitive forward design methods work for simple optical elements, they often fail to produce complicated elements, especially those involving multiple elements. We present an inverse design methodology for large arrays of wavelength scale spheres based on both adjoint optimization or sensitivity analysis and generalized multi-sphere Mie theory as a solution to the design problem. We validate our methodology by designing two sets of optical elements with scatterers on sub-wavelength and super-wavelength periodicity grids. Both sets consist of a singlet and a doublet lens, with one and two layers of spheres respectively designed for 1550 nm. The designed NA is ~0.33 (~0.5) for the sub-wavelength (super-wavelength) periodic structure. We find that with the sub-wavelength periodicity, the full width at half maximum of the focal spot produced by the singlet and doublet is smaller than that produced by an ideal lens with the same geometric parameters. Finally, we simulate a realistic experimental scenario for the doublet where the spheres are placed on a substrate with the same refractive index. We find the performance is similar, but with lower intensity at the focal spot and larger spot size. The method described here will simplify the design procedure for complicated multifunctional optical elements and or scatterer array-based volume optics based on a specified figure of merit.**

## 1. Introduction

Research on arrays of wavelength scale dielectric optical scatterers has resulted in micrometer scale diffractive optical elements capable of both imitating existing refractive optical elements [1-9], and also performing new optical functions [3, 10-12]. These devices represent the current state of the art in ultra-thin, high quality optical elements. This wealth of research has generally been performed using forward design methods, with notable exceptions[9, 13], and there is great interest in an inverse design method capable of producing a fully three dimensional optical element. In forward design methods, we calculate a spatially varying phase profile to implement a desired optical function; we then choose from a library of electromagnetic scatterers with precomputed complex amplitude coefficients in reflection or transmission to implement the calculated phase profile. A forward design approach is sufficient for optical functions with well-defined analytical forms such as those for a conventional lens, a vortex beam plate or higher–order polynomial phase-plates.

However, the forward design method can only explore a small sub-set of the whole gamut of optical elements, that can be built by controlling and fine-tuning the scattering of electromagnetic waves at the single scatterer level. We believe that the greatest potential of the scatter array based devices lies in such ability to engineer single scatterer. In a typical compact optical element, the number of scatterers may be in the thousands, if not many more. These large systems of scatterers have a large design space characterized by variables including the geometric and material properties of the scatterers themselves in addition to their spacing. In such systems, the sheer number of design variables precludes the possibility of efficiently fine-tuning individual scatterers using a forward design method. Forward design methods also fall short in more extreme scenarios involving non-paraxial optics such as high-NA lenses[14] or for the design of complex and multi-functional optical elements[15, 16] where there is no convenient analytical description. Moreover, it becomes difficult to design optical elements involving multiple such scattering surfaces. An inverse electromagnetic design method based on a specified device performance provides an attractive solution.

To date, inverse electromagnetic design methods have been applied to phase profile design[17, 18], single scatterer design[13], beam steering[19-22], and achromatic metasurface optics[9]. Several optimization methods, including gradient-based methods, particle swarm optimization [19] and genetic algorithms[9, 22] have been applied in designing optical elements. Genetic and particle swarm optimizations make no assumptions about the system to be optimized allowing them to be used in a wide variety of applications, but they are generally not suitable for large sets of variables[23]. In contrast, gradient descent methods are more specialized and utilize the physics of the system to iteratively converge towards local optima [20, 24-29].

Adjoint optimization-based gradient descent methods have been explored more recently for use in nanophotonics with great success in designing plasmonic nanostructures[20] and dielectric photonic elements[20, 25, 26, 30-32]. The benefits of adjoint optimization (or adjoint sensitivity analysis) have already been recognized in the fields of structural and aerospace engineering [24, 27, 33]. Within the electromagnetics community, adjoint optimization was first adopted at microwave frequencies[34, 35], but it has more recently been applied at optical frequencies. These methods generally solve Maxwell's equations in a meshed design space, and allow the refractive index or permittivity distribution to vary at each mesh point[13, 26, 29-32]. Unfortunately, these methods do not scale well with larger systems with small feature sizes as their accuracy and speed depend strongly on the mesh size.

In this letter, we propose a gradient-based inverse design method for spherical arrays of scatters using adjoint optimization and generalized multi-sphere Mie theory (GMMT). Instead of solving Maxwell's equations directly, we choose to work within the framework of GMMT, a special case of the T-matrix formalism[36-38]. GMMT describes the scattering of monochromatic electromagnetic radiation from an aggregate of spheres in spherical coordinates[36, 37]. The scattered field is completely described in terms of complex coefficients of the spherical vector wave function (SVWF) expansion[36-38]. In doing this, we trade the flexibility of designing arbitrarily shaped scatterers for the analytical scattering expressions of spheres to reduce computational costs. Specifically, this allows us to avoid meshing the individual scatterers that compose our device. Rather than optimizing over a refractive index distribution, we optimize over the radii of the spheres to achieve our figure of merit (FOM). For this method, the iteration time scales with the total number of spheres, their density, and the field expansion order, rather than the total device size[38].

## 2. Generalized Multi-sphere Mie Theory

We will briefly summarize the relevant aspects of GMMT here. A more detailed description of the formalism of GMMT can be found elsewhere[36-38]. Following the notation and work of Egel et al.[38], we begin with an array of $N$ homogenous, separated spheres $S_i$, where each sphere is characterized by its center position $r_i$, radius $R_i$, and refractive index $n_i$, where $i = 1,2,\ldots N$. The spheres are embedded in a background refractive index $n_0$ and illuminated by a monochromatic plane wave, $E_{in}(r)$ with the implicit $\exp(-i\omega t)$ dependence suppressed.

### A. Single sphere

In the case of scattering from a single sphere, our problem is exactly that of the Mie solution. Taking a single sphere $S_i$, we can write the total electric field as a sum of the known incoming field and the scattered field. These are expanded in spherical coordinates using the regular and outgoing SVWFs.

$$E(r) = E_{in}^i(r) + E_{scat}^i(r), (1)$$

with

$$E_{in}^i(r) = \sum_n a_n^i \psi_n^{(3)}(r - r_i), (2)$$
$$E_{scat}^i(r) = \sum_n b_n^i \psi_n^{(3)}(r - r_i). (3)$$

Here $a_n^i$ and $b_n^i$ are the SVWF coefficients of the incoming and scattered field of the $i^{th}$ sphere, respectively; $n$ is a multi-index that includes the polarization (TE and TM) and multipole indices $l = 1,2,\ldots$ and $m = -l,\ldots 1,2,\ldots l$. $\psi_n^{(1)}$ and $\psi_n^{(3)}$ are the regular and outgoing SVWFs, respectively. These coefficients are related by the T-matrix, $T_{n\,n'}^{i\,i'}$ as

$$b_n^i = T_{n\,n'}^{i\,i'} a_{n'}^{i'}. (4)$$

For an isotropic sphere, the T-matrix is diagonal, lacks any $m$ dependence, and is populated by the Mie coefficients obtained by solving the single sphere scattering problem.

### B. Multiple spheres

In the case of multiple spheres, the field incident on each sphere $S_i$ is not only the field from the original incident wave, but also the scattered field from other spheres described by:

$$E_{in}^i(r) = E_{in}(r) + \sum_{i' \neq i} E_{scat}^{i'}(r), (5)$$

This results in a system of coupled linear equations for the scattering coefficients $b_n^i$:

$$M_{n\,n'}^{i\,i'} b_{n'}^{i'} = T_{n\,n'}^{i\,i'} a_{in,n'}^{i'} (6)$$
$$M_{n\,n'}^{i\,i'} = \delta_{i\,i'} \delta_{n\,n'} - T_{n\,n''}^{i\,i'} W_{n''\,n'}^{i''\,i'}, (7)$$

where $a_{in,n'}^{i'}$ represent the coefficients corresponding to the incident field, and $W_{n''\,n'}^{i''\,i'}$ is the coupling matrix that relates the $i'$th sphere's scattered field to that of the $i''$th sphere incident field. This matrix depends only upon the positions of the spheres, and its form is further elaborated on in reference [38].

We solve this forward problem for the scattering coefficients $b_n^i$ using CELES, an open source CUDA-accelerated MATLAB package[38]. The parallelism offered by CELES allows us to quickly solve large systems of equations with thousands of spheres, and this robust forward solver is essential for the inverse design method.

## 3. Adjoint Optimization

The inverse design method iteratively changes an initial array of spheres to optimize a FOM describing the optical element's performance. We begin with a preset grid of sphere locations, and initialize spheres to be of the same radii at all points on the grid. During our inverse design process, the individual sphere radii serve as our only modifiable parameters. For a system of $N$ spheres, we have $N$ variables. To calculate the gradient, we use the adjoint method, which is well described in other works[20, 26, 30, 31] so we only cover formulating our specific problem here.

### A. Calculating the gradient

Given a FOM $f(b(R), R)$ where $b$ is a vector containing the scattering coefficients $b_n^i$, and $R$ is a vector containing the individual sphere radii, we are interested in the gradient of the FOM with respect to the parameters, $R$:

$$\nabla_R f = \left[\frac{\partial f}{\partial R_1}, \frac{\partial f}{\partial R_2}, \ldots, \frac{\partial f}{\partial R_N}\right]. (8)$$

Taking a specific sphere radius $R_j$, and using the chain rule, we can write:

$$\frac{\partial f}{\partial R_j} = 2\, Re\left\{\frac{\partial f}{\partial b}\frac{\partial b}{\partial R_j}\right\}, (9)$$

where the first term is easily calculated, but the second term is more troublesome. We use the Wirtinger derivative to calculate the first term due to the complex nature of $\boldsymbol{b}$[30]. As the expression does not guarantee a strictly real gradient, we only choose the real part[26]. The main advantage of the adjoint method is it allows us to avoid explicitly calculating the second term using algebraic manipulations and matrix associativity. We can take a derivative of equation (6) with respect to a specific sphere radius $R_j$, and recognizing that $W_{n''n'}^{i''i'}$ is independent of particle radii, we obtain:

$$M_{nn'}^{ii'}\frac{\partial b_{n'}^{i'}}{\partial R_j} = \frac{\partial T_{nn'}^{ii'}}{\partial R_j} a_{in,n'}^{i'} + \frac{\partial T_{nn''}^{ii''}}{\partial R_j} W_{n''n'}^{i''i'} b_{n'}^{i'}. \quad (10)$$

We can solve (10) for $\frac{\partial b_{n'}^{i'}}{\partial R_j}$ by inverting $M$. Then inserting it into (9) gives us:

$$\frac{\partial f}{\partial R_j} = \frac{\partial f}{\partial b_{n'}^{i'}}\left(M_{nn'}^{ii'}\right)^{-1}\left(\frac{\partial T_{nn'}^{ii'}}{\partial R_j} a_{in,n'}^{i'} + \frac{\partial T_{nn''}^{ii''}}{\partial R_j} W_{n''n'}^{i''i'} b_{n'}^{i'}\right). \quad (11)$$

From (11) we can define a new variable $\lambda_n^i$, the 'adjoint' coefficients from the first two terms on the right side, and obtain a second or 'adjoint' system of equations. The adjoint coefficients and the derivatives of the Mie coefficients in $T_{nn'}^{ii'}$ are the only new quantities we must calculate. We populate $\frac{\partial T_{nn'}^{ii'}}{\partial R_j}$ with expressions of Mie coefficient derivatives found in literature[39, 40]. The adjoint system of equations is given by:

$$\left(\lambda_n^i\right)^T = \frac{\partial f}{\partial b_{n'}^{i'}}\left(M_{nn'}^{ii'}\right)^{-1} \Rightarrow \left(M_{nn'}^{ii'}\right)^T \lambda_n^i = \left(\frac{\partial f}{\partial b_{n'}^{i'}}\right)^T. \quad (12)$$

In this form, we see $\lambda_n^i$ is independent of the sphere radius $R$, with respect to which we are taking a derivative. Hence, it can be computed once and stored. This allows us to compute the full gradient by solving (12) for $\lambda_n^i$ and performing $N$ matrix multiplications according to:

$$\frac{\partial f}{\partial R_j} = 2\, Re\left\{\left(\lambda_n^i\right)^T\left(\frac{\partial T_{nn'}^{ii'}}{\partial R_j} a_{in,n'}^{i'} + \frac{\partial T_{nn''}^{ii''}}{\partial R_j} W_{n''n'}^{i''i'} b_{n'}^{i'}\right)\right\}, \quad (13)$$

where previously we needed to solve $N$ systems of equations. While our iteration time still ultimately depends on the numbers of spheres in our system, we have removed the explicit dependence on the number of variables.

We begin with an initial condition of identical spheres on a square grid as shown in figure 1a. In our optimization process, we alternate between solving the forward problem for computing the FOM and the adjoint problem for computing the gradient (figure 2b).

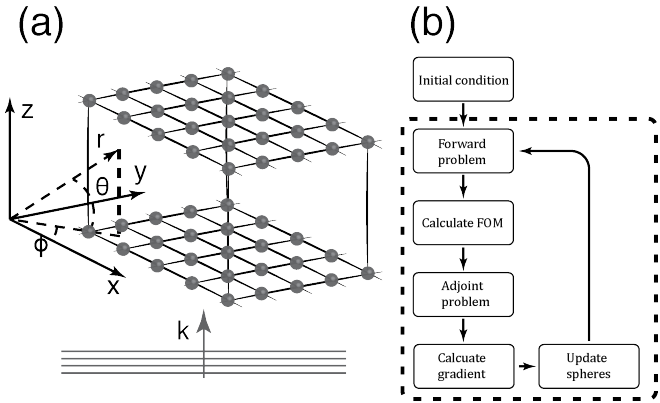

Figure 1: Optimization setup and flow. (a) Example schematic of an initial condition for our optimization process. Pictured is an array of identical spheres consisting of two layers of spheres arranged on a square grid with a plane wave of wave vector k incident from below. Cartesian (spherical) axes are in solid (dotted) lines. The optical axis is along z. (b) Process flow of an optimization process. The steps inside the dotted box are those executed in a single iteration of the inverse design method.

The radii of the array of spheres are continually updated using gradient descent with a fixed step size as shown in figure 1b. The optimization routine runs up to a fixed number of iterations, in our case chosen to be 200. All of the output radius distributions from the optimization process are then simulated using Lumerical FDTD Solutions.

**B. Figure of Merit**

So far, we have explained the method of obtaining the gradient of the FOM. We now focus on the FOM itself. In the following we will describe the optimization of intensity at a single point. The impact of the FOM is coded into the adjoint system of equations by the term $\frac{\partial f}{\partial \boldsymbol{b}}$. It is best if the FOM is smooth and has an explicit dependence on our scattering coefficients $\boldsymbol{b}$ to make computation of the derivative simple and cheap.

We consider maximizing the intensity at a single focal point $\boldsymbol{r_0}$ in space given by:

$$f(\boldsymbol{b},\boldsymbol{R},\boldsymbol{r_0}) = |\boldsymbol{E}(\boldsymbol{b},\boldsymbol{R},\boldsymbol{r_0})|^2 = \left|\sum_{i,n} b_n^i(\boldsymbol{R})\boldsymbol{\psi}_n(\boldsymbol{r_0})\right|^2. \quad (14)$$

We are primarily interested in the quantity $\frac{\partial f}{\partial b_n^i}$. As we are differentiating with respect to a complex variable, we use the Wirtinger derivative, and our final expression becomes:

$$\frac{\partial f}{\partial b_n^i} = conj\left(b_n^i(\boldsymbol{R})\boldsymbol{\psi}_n(\boldsymbol{r_0})\right) \cdot \boldsymbol{\psi}_n(\boldsymbol{r_0}), \quad (15)$$

where $conj()$ denotes the complex conjugate. These adjoint coefficients are computed and stored after every iteration.

## 4. Results

In this section, we present two sets of optical elements designed using the optimization method described earlier. Each set consists of a singlet and a doublet consisting of one and two layers of spheres, respectively. We choose radii and periodicity such that there is no overlap between adjacent spheres, as our implementation of GMMT does not currently support overlapping structures. All the elements are designed to maximize the intensity at a point 50 μm away from the center of the lens. The spheres have a refractive index of 1.52, corresponding to that of the highest resolution resist available for a state-of-the-art commercial two-photon lithography system (Photonic Professional GT, Nanoscribe GmbH, Germany). The plane of the lens is taken to lie in the x-y plane, with the optical axis along z, perpendicular to the plane of the lens. Our excitation is a monochromatic plane wave polarized along the y-axis at normal incidence.

For all optimizations, we assume the spheres are suspended in vacuum, and do not include a substrate as our implementation of GMMT does not readily support closely packed non-spherical particles. GMMT may be extended with the use of coordinate transformations to support the effect of a substrate on an aggregate of spheres[41]. We choose to cut off our field expansions at multipole order $l$ = 3 for the sub-wavelength periodicity, and $l$ = 4 for the super-wavelength periodicity. This cutoff is based on the scattering properties of our spheres, and the justification is presented in Appendix B.

### A. Sub-wavelength periodicity

Here we present two optical elements designed with a sub-wavelength periodicity. Scatterer-based optical elements with sub-wavelength periodicity, also known as metasurfaces, have been of great interest due to their ability to control the propagation of diffraction orders. For both elements, we begin with a square grid of periodicity 1240 nm populated with identical spheres of radius 300 nm. We allow the sphere radii to vary continuously between 150 nm and 600 nm, making our multipole cutoff $l = 3$. For further discussion on the dependence of the multipole cut off on the sphere parameters, we refer to Appendix B.

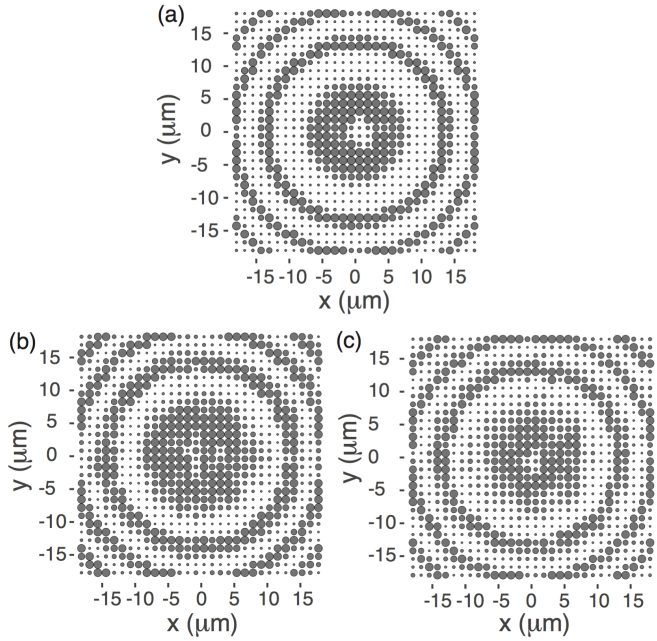

Figure 2: Final radius distribution of the sub-wavelength elements with periodicity 1240 nm. Radii are allowed to range from 150 nm to 600 nm. (a) sub-wavelength singlet. (b), (c) sub-wavelength doublet top and bottom layers respectively.

For the singlet lens, we begin with a single layer of 30×30 of spheres resulting in final dimensions of $36 \mu m \times 36 \mu m \times 1.2 \mu m$. The final radius distribution of the optimization process is shown in figure 2a. We can see that the result is mostly circularly symmetric as expected for a lens. There is some asymmetry near the origin of the lens. We attribute this to our algorithm running for a fixed number of iterations rather than convergence at a local optimum. Our gradient-based method only guarantees convergence to a local optimum, and we manually terminate the algorithm when we achieve the desired performance as described by the FOM.

For the doublet lens, we begin with two layers of 30 x 30 spheres, separated by 2 μm center-to-center distance resulting in final dimensions of $36 \mu m \times 36 \mu m \times 4 \mu m$. The final radius distribution of the optimization process is shown in figures 2b, c. We see again that the result is mostly circularly symmetric, with some asymmetry near the center of the lens.

*1. Singlet*

We simulate the singlet output radius distribution in FDTD and see a clear focal spot at the design focus of 50 μm in the x-z plane (figure 3a). In addition, we characterized the performance under chromatic illumination and see that the focal length changes linearly with wavelength (figure 3b). The focal length of the singlet shifts by ∼7 μm over the 200 nm illumination bandwidth. This dependence is consistent with that of conventional diffractive optics. Lastly, we characterize the spot size produced by the lens. We calculate the full width at half maximum (FWHM) of the focal spot at 50 μm for different wavelengths by fitting 1-D slices (x and y) of the intensity peak to Gaussians (figure 3c). The FWHM along the x and y directions behave differently, corresponding to an asymmetric focal spot. We attribute this to the choice of our FOM, which does not constrain the optimization routine to a symmetric focal spot. We compare these values to the FWHM of a diffraction-limited Airy disk corresponding to an ideal lens of the same geometric parameters and find that the singlet produces a smaller spot size. Examples of the fitting process and more details on the calculation of diffraction-limited FWHM are presented in Appendix C. The performance of the singlet when illuminated with an orthogonal linear polarization to the design linear polarization is presented in Appendix D.

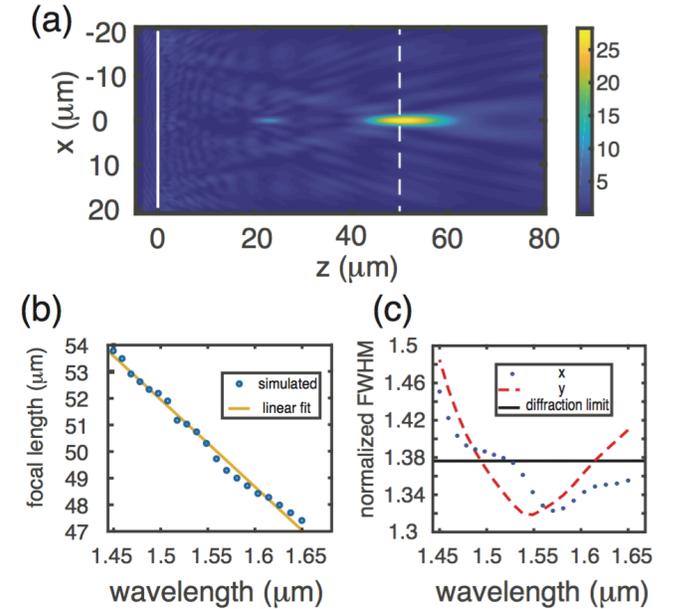

Figure 3: Performance of the sub-wavelength singlet in FDTD. (a) Intensity plot of the x-z plane under illumination at 1548 nm showing a clear focal spot at 50 μm. Data used to compute the FWHM is taken from slices in the plane of the white dashed line. The spheres are located at the solid white line. (b) Dependence of the focal length on incident wavelength, showing a linear dependence within this bandwidth. The focal length shifts 7 μm over this bandwidth. (d) Calculated FWHM of the singlet from data taken at the white dashed line in (a). The FWHM is obtained by fitting the intensity peak to a Gaussian. Blue dotted and red dashed lines represent fits from data taken along the x and y axes respectively. The black line is the diffraction limit for an ideal lens with the same geometric parameters as the design. All FWHM are normalized by their respective wavelengths.

*2. Doublet*

To demonstrate the suitability of the algorithm for designing fully three-dimensional arrays of scatters, we also demonstrate a doublet design with two layers of spheres. We simulate the output radius distribution in FDTD and see a clear focal spot at the design focal length of 50 μm in the x-z and y-z planes (figure 4a). We do the same set of characterizations as for the singlet lens, and find that the focal length shift is almost the same for the doublet design (figure 4b). However, we see a noticeable improvement in the spot size at the focal length compared to that of the singlet (figure 4c). The FWHM decreases across the entire bandwidth with the additional layer of spheres. Again, the FWHM along the x and y directions show asymmetric behavior, and we attribute this to our FOM.

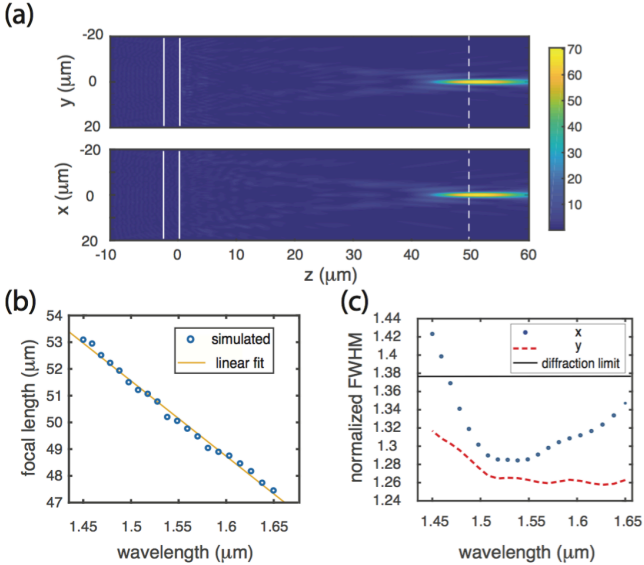

Figure 4: Performance of the sub-wavelength doublet in FDTD. (a) Intensity plot of the x-z and y-z plane showing clear focusing at the design focal length of 50 μm under illumination by 1548 nm. Data used to compute the FWHM is taken from the white dashed line, and the layers of spheres are located at the solid white lines. (b) Dependence of the doublet focal length on illumination wavelength. The line is a linear fit to the data. (c) Calculated FWHM using data from the white dashed line in (a). The FWHM is obtained by fitting the intensity peak to a Gaussian. Blue dotted and red dashed represent fits from data taken along the x and y axes respectively. The black line is the diffraction limit for an ideal lens with the same geometric parameters as the design. All FWHM are normalized by their respective wavelength

**B. Super-wavelength periodicity**

We now present two optical elements designed with a super-wavelength periodicity. In contrast to the sub-wavelength periodic devices, these devices have dimensions that can exceed the wavelength of incident light. As such, they represent devices that can be more easily fabricated with available fabrication techniques. For both elements, we begin with a square grid of periodicity 2050 nm populated with identical spheres of radius 700 nm. We allow the sphere radii to vary continuously between 150 nm and 1000 nm, making our multipole cutoff $l = 4$. For further discussion on the multipole cutoff, we refer to Appendix B.

For the singlet lens, we begin with a 30×30 layer of a spheres with final dimensions of $60 \mu m \times 60 \mu m \times 2 \mu m$. The final radius distribution is shown in figure 5a.

For the doublet lens, we begin with two 30×30 layers of spheres, separated by a center-to-center distance of 4 μm. This results in final dimensions of $60 \mu m \times 60 \mu m \times 8 \mu m$. The top and bottom layer are shown in figures 5b, and 5c respectively.

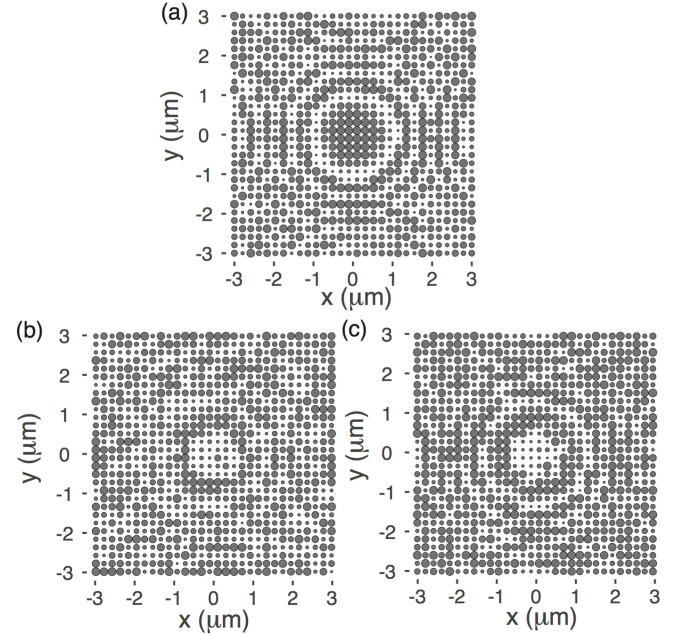

Figure 5: Final radius distribution of the super-wavelength elements with periodicity 2050 nm. Radii are allowed to range from 150 nm to 1000 nm. (a) super-wavelength singlet. (b), (c) super-wavelength doublet top and bottom layers respectively.

*1. Singlet*

From our simulation, we find a clear focal spot at a focal length of 45 μm in both the x-z and y-z planes for illumination under 1548 nm illumination (figure 6a). This result is not consistent with the design focal length of 50 μm at 1550 nm. We attribute this difference to the meshing of our design space. As the super-wavelength periodic devices are larger than the sub-wavelength devices, we were unable to mesh the super-wavelength devices to the same accuracy. Again, this device's focal length displays a linear dependence on wavelength in this bandwidth (figure 6b). Lastly, we calculate the FWHM at the focal length of 45 μm and find spot sizes on the order of the diffraction-limited FWHM (figure 6c), with no spot sizes smaller than the diffraction-limited FWHM. The diffraction limit is calculated for a lens with the same geometric parameters, and a focal length of 45 μm.

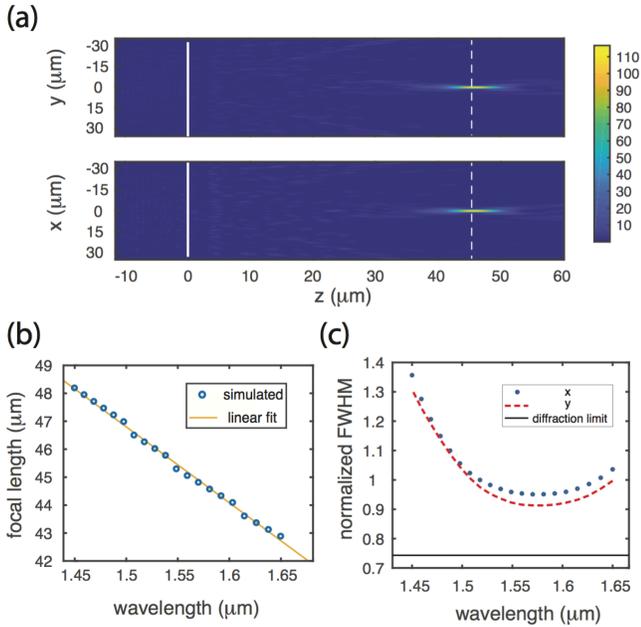

Figure 6: Performance of the super-wavelength singlet in FDTD. (a) Intensity plot of the x-z and y-z planes showing clear focusing at 45 μm under illumination by 1548 nm light. Data used to compute the FWHM is taken from the white dashed line, and the layer of spheres is located at the solid white line. (b) Dependence of the focal length on illumination wavelength showing a clear linear dependence in this bandwidth. (c) Calculated FWHM using data from the white dashed line in (a). The FWHM is obtained by fitting the intensity peak to a Gaussian. Blue dotted and red dashed lines represent fits from data taken along the x and y axes respectively. The black line is the diffraction limit for an ideal lens with the same geometric parameters as the design. All FWHM are normalized by their respective wavelength

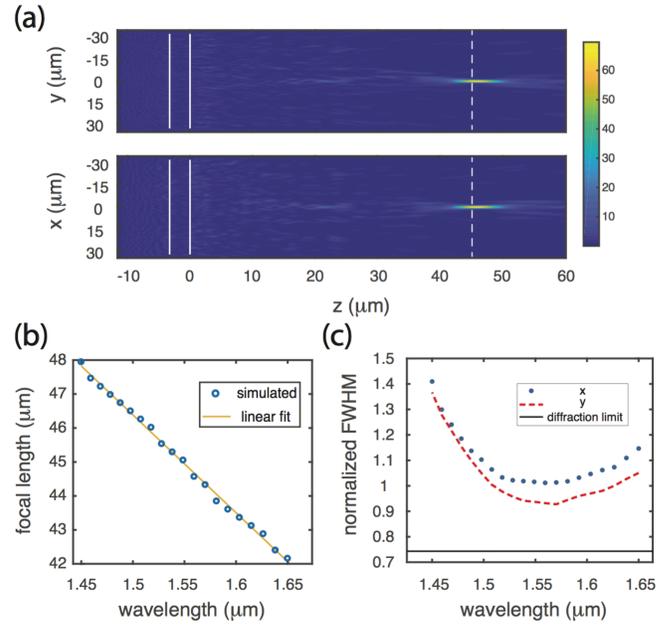

Figure 7: Performance of the super-wavelength doublet in FDTD. (a) Intensity plot of the x-z and y-z planes showing clear focusing at 45 μm under illumination by 1548 nm light. Data used to compute the FWHM is taken from the white dashed line, and the layers of spheres are located at the solid white lines. (b) Dependence of the focal length on illumination wavelength showing a clear linear dependence in this bandwidth. (c) Calculated FWHM using data from the white dashed line in (a). The FWHM is obtained by fitting the intensity peak to a Gaussian. Blue dotted and red dashed lines represent fits from data taken along the x and y axes respectively. The black line is the diffraction limit for an ideal lens with the same geometric parameters as the design. All FWHM are normalized by their respective wavelength

*2. Doublet*

In our simulation we find a clear focal spot at 45 μm in both the x-z and y-z plane (figure 7a). This is not the design focal length, and as with the super-wavelength singlet, we attribute this difference to the discretization of our mesh. This device also displays the same dependence on incident wavelength within this bandwidth as the singlet (figure 7b). The FWHM is similar to that of the super-wavelength singlet, though generally the doublet displays a larger spot size (figure 7c). We currently do not understand why the super-wavelength doublet does not result in better focusing performance. We would expect the addition of a second layer of spheres to improve the performance of the device. The diffraction limit is calculated for a lens with the same geometric parameters, and a focal length of 45 μm.

### C. Substrate effects

In any experimental demonstration of the elements presented, it is likely that the spheres will be sitting on a substrate. Hence, we simulate the sub-wavelength doublet in FDTD after adding two 800 nm spacer layers, one between the top and bottom layers, and one below the bottom layer with the same refractive index as the spheres (figure 8). The bottom of the simulation region is then filled with a quartz substrate ($n$ = 1.45).

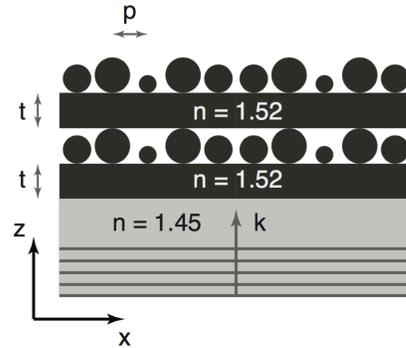

Figure 8: Substrate schematic showing the doublet with the spacer layers and substrate. The thickness of the spacer layers (t) is 800 nm, periodicity (p) is 1240 nm, and light with wave-vector k is incident from below through a quartz substrate $n$ = 1.45.

In our simulation, we find a clear focal spot at the design focal length of 50 μm in both the x-z and y-z plane (figure 9a). However, the intensity at the focal spot shows a noticeable decrease. The focal length displays a linear dependence on illumination wavelength in this bandwidth (figure 9b). The spot sizes produced by the device are larger than those of the ideal device, but there is still a large range of wavelengths where the FWHM is smaller than the FWHM of a diffraction-limited lens (figure 9c). We see that the displacement of the spheres, addition of the two spacer layers, and addition of a substrate have a noticeable effect on the performance of the device. However, the focal length remains unchanged, and the focal spots are of similar size.

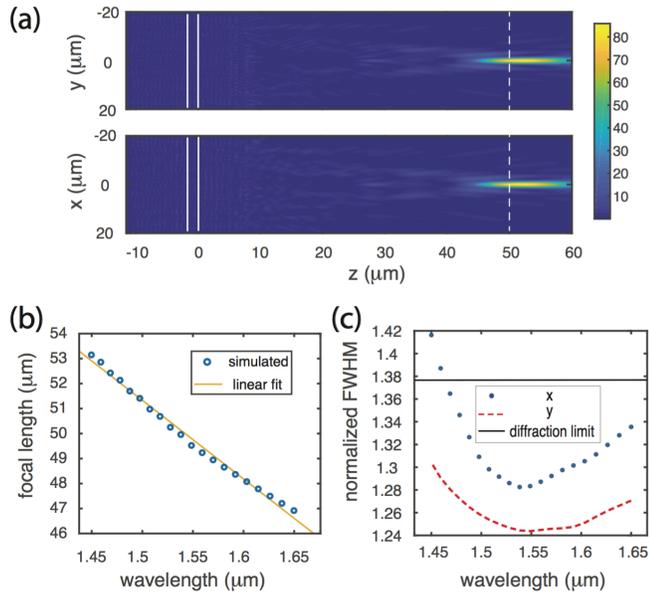

Figure 9: Performance of sub-wavelength doublet with added dielectric substrate layers in FDTD. (a) Intensity plot of the x-z and y-z plane showing clear focusing at the design focal length of 50 μm under illumination by 1548 nm. Data used to compute the FWHM is taken from the white dashed line, and the layers of spheres are located at the solid white lines. (b) Dependence of the doublet focal length on illumination wavelength showing a linear dependence. (c) Calculated FWHM using data from the white dashed line in (a). The FWHM is obtained by fitting the intensity peak to a Gaussian. Blue dotted and red dashed lines represent fits from data taken along the x and y axes respectively. The black line is the diffraction limit for an ideal lens with the same geometric parameters as the design. All FWHM are normalized by their respective wavelength

## 5. Discussion

We have designed and simulated two sets of simple optical components using an inverse design method based on adjoint optimization and generalized multi-sphere Mie theory. Specifically, we designed a singlet and doublet on a sub-wavelength periodicity grid with NA ~ 0.5, and a singlet and doublet on a super-wavelength periodicity grid with NA ~ 0.33. The performance of the devices has been checked for accuracy using FDTD. All of the devices are designed by simply specifying an initial condition and desired figure of merit. The devices display chromatic aberrations consistent with diffractive optics.

In addition, we presented simulations showing the ability of the sub-wavelength periodicity doublet to perform when combined with a substrate and spacing layers. Under ideal conditions, and also with the substrate, the sub-wavelength doublet was able to focus to a spot with FWHM smaller than the diffraction-limited FWHM in the far field. This method will be most useful in designing complex optics such as high-NA lenses and volume optics, when using physical intuition to guide forward design methods is unsatisfactory. Systems with sub-wavelength resolution have been demonstrated in the past at both the near field[42, 43] and far field[44-46]. The parameters we have chosen are at the very limit of current two-photon optical lithography systems. Researchers have proven two-photon lithography is able to produce high quality, complex optical elements such as compact objectives[47] and freeform elements[48] for visible frequencies, and thus the simulated devices can also be fabricated.

## 6. Conclusion

We describe an inverse design method using adjoint optimization and GMMT. We have demonstrated the suitability of this method for designing fully 3-dimensional single and multi-layer optics by presenting the performance of two sets of singlet and doublet lenses. While we have chosen to present results within the infrared regime, the method makes no assumptions about the wavelength of light, and is well-suited for calculating scattering from any wavelength scale distribution of spheres. This work constitutes a significant step forward in the use of inverse design techniques to design scatterer array-based optical elements.

## 7. Appendices

### A. Machine specifications

CentOS 7
MATLAB v9.2 r2017a with Parallel Computing Toolbox v2.4
2x Intel E5-2620 @2.1 GHz
NVIDIA Tesla K40 12 GB Memory running CUDA 9
64 GB DDR3 Memory

It is worthwhile to note that while our computer has 12 physical cores, our optimization process generally relies only on a single CPU thread barring computation of the block diagonal preconditioner. Intensive parallel tasks are instead passed to the GPU for computation whenever possible.

Our inverse and forward design problems are solved using a modified version of CELES, and all of our FDTD simulations are performed using Lumerical FDTD Solutions. More details on the performance of CELES is available from Egel et al[38].

CELES is available free of charge, and our implementation of the optimization algorithm is available on request.

### B. Expansion order cutoff

The iteration time of the inverse design method depends on both the particle number and the expansion order. Larger numbers of particles or expansion orders increase the iteration time. As we are interested in very large arrays of particles, it is important to find a reasonable cutoff expansion to balance the speed of the iteration, and the accuracy of the result.

The valid cutoff expansion order is ultimately determined by the scattering properties of individual spheres described by their Mie coefficients. These scattering properties are determined by the geometric and material properties of the sphere in addition to the wavelength of input light. Figure 10 shows the absolute value of the calculated Mie coefficients for our spheres with radii ranging from 150 to 1500 nm, and refractive index 1.52 under illumination by 1550 nm input. The dashed boxes show the geometric parameters we use. We see that these spheres we use in the sub-wavelength regime respond strongest to spherical waves up to $l$ = 3, justifying our cutoff. For the super-wavelength regime, there is some contribution from $l$ = 4 from the largest spheres, so we expand further than for the sub-wavelength regime

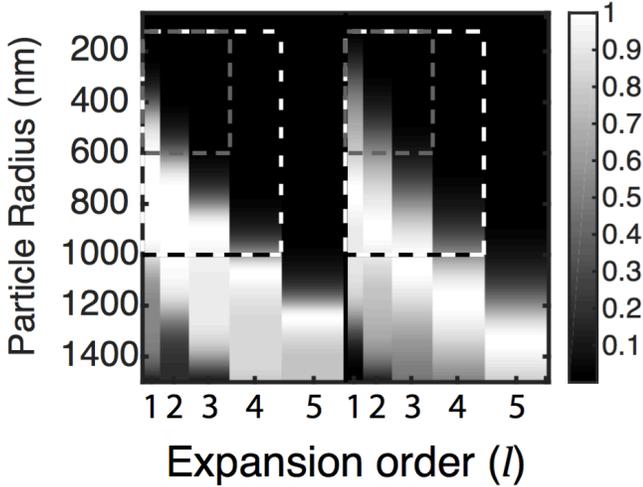

Figure 10: Absolute value of Mie coefficients under illumination by 1550 nm light. Light (dark) dashed boxes indicate the range of parameters used for the designs in the sub-wavelength (super-wavelength) periodicity section. The Mie coefficients for isotropic, uniform spheres have no $m$ dependence, so all coefficients of given expansion order $l$ are equivalent. The two sets of data correspond to two polarizations.

From the Mie coefficients, we can find the optical response of particles of varying radii. It is clear that the use of larger spheres requires larger expansion orders, increasing computational time. However, larger expansion orders also correspond to being able to control higher order spherical waves.

### C. Fitting the diffraction limit

We calculate the diffraction-limited FWHM by fitting an Airy disk corresponding to our choice of geometric parameters to a Gaussian. The intensity of an Airy disk is described by:

$$I(\theta) = I_o \left(\frac{2J_1(ka \sin \theta)}{ka \sin \theta}\right)^2, \quad (16)$$

where $J_1$ is the Bessel function of the first kind, $k$ is the wave vector of incident light in the medium immersing the lens, $a$ is the radius of the lens, and $\theta$ is the angular position from the focal point. Figure 11a is an example of an Airy disk calculated for our geometric parameters of a lens radius 18 μm, and focal length 50 μm under illumination by 1548 nm light. Figure 11b and 11c are the x and y fits to simulation data for the doublet lens under the same illumination. We can see the side lobes for the doublet are much more pronounced than that of the ideal lens. It is known that spots with sub-diffraction limited FWHM can be generated by diverting power from the central maximum into the side lobes[49].

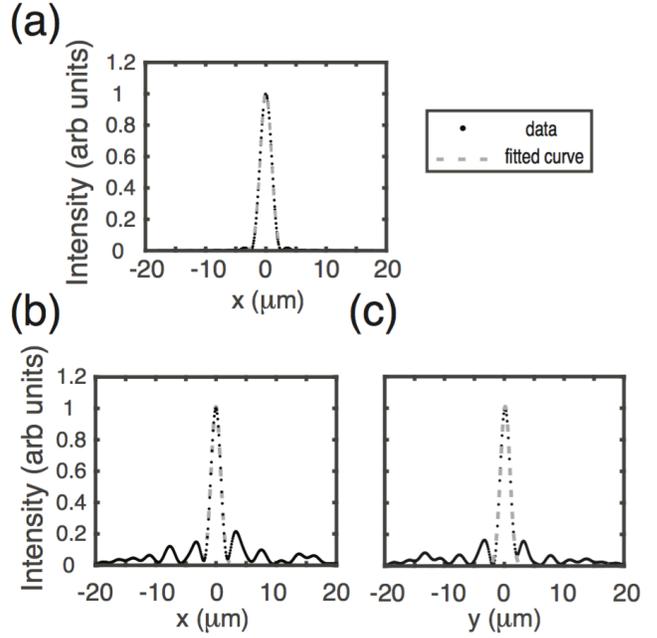

Figure 11: Example fits for an ideal lens and our simulated design. (a) is the Airy disk generated by an ideal lens under illumination by 1548 nm with radius 18 μm, and focal length 50 μm. (b) and (c) are fits along the x and y axes respectively for the doublet design presented in the main text under 1510 nm illumination.

### D. Polarization effects

In our optimization process, we assume a single linear polarization (the design polarization), while neglecting to optimize the performance for the orthogonal polarization. We simulate the performance of the sub-wavelength singlet under illumination by the orthogonal polarization. The design still focuses to a similar intensity, and produces a similar field profile to that of the design polarization. We plot the dependence of the focal length on illumination wavelength and find is the same as with illumination by the design polarization (figure 12a). The spot size along the x direction shows a notable increase when illuminated with the orthogonal polarization (figure 12b). However, the spot size along the y direction remains similar. The increase in FWHM along the x direction can be attributed to the change in polarization of the incident light.

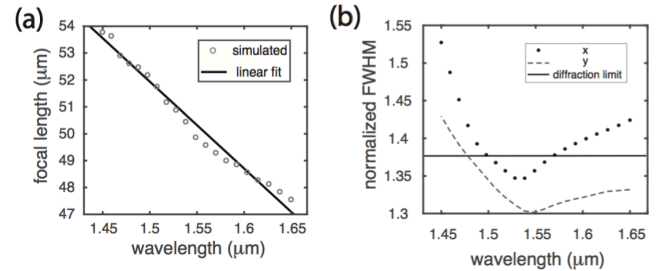

Figure 12: Polarization dependence of the sub-wavelength singlet. (a) Focal length dependence on wavelength, and (b) spot size (FWHM) dependence on wavelength calculated at 50 μm. Solid black line is the calculated diffraction-limited FWHM, black dots (dotted grey line) is the FWHM along the x (y) direction.

**Funding sources and acknowledgments.** Funding of this research is provided by the startup fund from University of Washington, Seattle.

**Acknowledgment**. We thank NVIDIA corporation for the contribution of the GPU for our calculations. In addition, we thank Amos Egel, Lorenzo Pattelli, and Giacomo Mazzamuto for allowing us to use and further contribute to CELES.